\documentstyle[12pt]{article}

\begin{document}

\def\eq{\begin{equation}}
\def\en{\end{equation}}

\begin{titlepage}

\vskip .27in
\begin{center}
{\large \bf CONDUCTANCE STATISTICS IN SMALL GaAs:Si WIRES AT LOW
TEMPERATURES.
\par
I. THEORETICAL ANALYSIS: TRUNCATED QUANTUM FLUCTUATIONS IN INSULATING WIRES.}

\vskip .27in
Francois Ladieu$^{*}$ and Jean-Philippe Bouchaud$^{*}$
\vskip .2in
{\it
* Service de Physique de l'Etat Condens{\'e}, CEA-Saclay, 91191 Gif s/ Yvette
CEDEX}

\end{center}
   \par
\vskip .2in
\begin{center} { \bf Abstract } \end{center}
\vskip .8in
We analyse in detail Mott's variable range hopping in one dimension,
expanding on earlier work by Raikh
and Ruzin. We show that the large conductance
fluctuations in disordered insulators result from a subtle
interplay between purely quantum phenomena and geometrical fluctuations arising
from the energies and locations
of the impurities. Our results compare very well with both
experiments
and numerical simulations.

\end{titlepage}
\pagebreak

In recent years, theoretical and experimental works have underlined the
importance of conductance fluctuations in disordered conductors \cite{Revue}.
While the
situation is rather well understood in the metallic phase, where `universal
conductance fluctuations' $\delta g \simeq {e^2\over \hbar}$ are observed,
less is known in the strongly localized regime. On the experimental side,
several groups have observed reproducible conductance fluctuations on quasi
one dimensional (insulating) wires \cite{IBM}, \cite{Sanquer}.
These fluctuations have been argued to
be of quantum origin \cite{Azbel} : the transmission coefficient of a
disordered bar is
known to depend sensitively on the energy (i.e. the gate voltage in the
experiments) and on the configuration of the impurities. The size of these
(zero temperature) fluctuations have been computed within the random matrix
theory,
\cite {Pichard} and the
result is that var $\ln g = -< \ln g > \equiv {L \over \xi}$, where $L$ is
the length of the sample and $\xi$ the localisation length. Another theory,
based on the `directed paths approximation' predicts somewhat smaller
fluctuations \cite{DP} : var $\ln g \propto C L^{2 \omega}$, where $\omega$ is
an exponent
close to $1/5$ (in three dimensions) and $C$ a prefactor independent of
$< \ln g >$: in other words, in this theory  var $\ln g$ and $< \ln g >$
may evolve differently with e.g. magnetic field or gate voltage.
For non zero temperatures, however, the inelastic collision time is much
shorter
than the time needed for the electron to cross the sample, and activated
energy jumps between impurities come into play: one enters Mott's
variable range hopping regime \cite{Mott}, \cite{Pollak}, where phase coherence
is only preserved at
length scales shorter than the typical jump size $r_0$. P. Lee \cite {Lee} has
argued, on the basis of numerical simulations (see also \cite{Serota}),
 that the `geometrical fluctuations' in the
location and energies of the impurities were sufficient, in the Mott regime,
to induce large, reproducible conductance fluctuations -
which would thus be unrelated to quantum, i.e. interference, effects.
Comparaison with the experiments, however, show that the size of
the fluctuations obtained in ref. \cite{Lee} is
too large \cite{Sanquer}. \par
The aim of the letter is twofold. We first show how
to solve analytically Mott's problem in one dimension. This provides a
clear interpretation of the conductance of a finite disordered wire, which is
limited by the `weakest link' in the sample, and allows to characterize the
fluctuations and discuss their dependence on the length of the sample.
These results are however almost identical to those of Raikh and
Ruzin \cite{Raikh} - although these authors relied on
a simplified analysis. As already discussed \cite{Raikh}, one
reaches
 quantitative
agreement with the numerical data \cite{Lee},\cite{Serota}
. To be able to describe all the features
of the experiments, we then show that quantum fluctuations governing
the fluctuations of the weakest link's conductivity must be taken into account.
However, due to the presence of other links,
these fluctuations cannot develop fully - they are `truncated' by geometrical
effects. Details concerning the experiments are published in the
companion
 paper \cite{Sanquer}.
\par
We thus start with Mott's problem in one dimension. We assume that the energy
levels are independent and randomly scattered with a density of states
 $\rho$ per unit
length and energy. The resistance between two sites $(x_i,E_i)$, $(x_j,E_j)$ is
given by (see \cite{Halp})

$$ R_{ij} = R^* \exp \left[{|E_i|+|E_j|+|E_i-E_j| \over 2kT} + {|r_i-r_j| \over
\xi}
\right] \eqno(1) $$

where $R^*$ is a typical (metallic) resistivity at scale $\xi$, which will be
set to $1$ in
the following, and $E=0$ corresponds
to the Fermi level. Note that eq. (1) completely neglects the fluctuations of
quantum origin, which can
be seen as giving a (gaussian) distribution to $1 \over \xi$ \cite{Pichard}.
These quantum fluctuations will
be considered later.
\par The problem is thus to determine from eq.(1) the total resistance
between $(x=0,E=0)$ and $(x=L,E=0)$ of a given sample, knowing the distribution
of
$(x_i,E_i)$. Even though the problem is one dimensional, this is not trivial
since each site
is connected to every other site. As usual for this problem \cite{Halp}, in
view of the exponentially fast
variation of $R_{ij}$, we approximate the end to end resistance
{\it with that of the less resistive path}. Furthermore, eq. (1) shows that the
`optimal'
path cannot wander arbitrarily far from the Fermi level. We shall thus make the
assumption
that {\it the optimal path can be constructed by always choosing locally the
least resistive link.}
 We will discuss later the validity of this assumption and show  that it leads
to the exact asymptotic
result. [It is, in any case, an upper bound for the resistance]. Now, suppose
that the electron
arrives on a site with energy $E \equiv mkT$. What is the probability
$\int_{n}^{\infty} dn P(n|m)$ that the least resistive
link from this site has a resistance greater or equal to $R \equiv e^n$ ? For
this to occur,
 no energy level must be found in the hatched area shown in fig. 1. This occurs
with probability \footnote
{ We neglect Wigner correlations between the energy levels, which is justified
as soon as $e^n \gg n_0^2$ \cite
{Imry}.}
$\propto \exp \left [-{n (n-m) \over n_0^2} \right ] $, where $n_0^2 \equiv {1
\over
\rho kT \xi}$: $\exp n_0$ is thus the usual Mott resistance, and $r_0 \equiv
n_0 {\xi \over 2}$ is the typical jump
size (which we assume to be much smaller than the sample length: we neglect the
possibility of resonant
tunneling).
 The total probability to find a given value
of $n \equiv \ln R$ is thus ${\cal P} (n) =
 \int_0^n dm P(n|m) {\cal Q}(m) =
 \int_0^n \left dm (2n-m) \over n_0^2 \right \exp \left [-{n (n-m) \over n_0^2}
\right ]
{\cal Q}(m)$. $\cal Q$ is the probability to find the electron at energy $m kT$
along the best
conducting path. [Note that it is {\it not} given by the Boltzmann weight !]. A
closed equation for $\cal P$
can be obtained by noticing that $n$ can
only increase when
$m$ increases, and hence $\int_n^\infty dn' P(n'|m)  =\int_0^m dm'
P(m'|n)$.
($P(m|n)$ is the probability to find the electron at energy $m$ after
crossing a resistance $e^n$). From this we obtain our central result for $\cal
P$:

$${ \cal P}(n) = \int_0^n \left dm  (2n-m) \over n_0^2 \right \exp \left [-{n
(n-m) \over n_0^2} \right ] \int_m^\infty
\left dn' n' \over n_0^2 \right \exp \left [-{n' (n'-m) \over n_0^2} \right ]
{\cal P} (n') \eqno(2) $$

This equation has been solved numerically : we show in fig. 1 (insert) ${\cal
P}$ versus the rescaled
variable $y={n \over n_0}$. One may show in particular that ${\cal}P(y) \simeq
y^2$ for small $y$, and
${\cal P}(y) \simeq a \exp [-{y^2 \over 2}] $ ($a \sim 2$) for large $y$
 \footnote {Although the whole calculation
only makes sense if $y < n_0$, beyond which $P(y) \equiv 0$}. This
asymptotic form of ${\cal P}$ coincides (apart from
the value of $a$) with the one obtained by Raikh
and Ruzin \cite{Raikh}, who in fact only computed
$\int_0^\infty dm P(n|m)$.
Before going further in our analysis, let us note that our hypothesis of
neglecting quantum fluctuations is only valid when the width of their
distribution $w_q$ is small enough compared to the width $w_{geo}$ of
 geometrical distribution $\cal P$ studied here. As shown by eq.(2),
 one has $w_{geo} \simeq n_0$, whereas one may estimate that
 $w_q \simeq n_0^ \omega$ ,with $\omega = {1 \over 2}$ in the barely
insulating regime and $\omega \sim {1 \over 5}$
 in the strongly localised regime. Therefore, our model will be valid when
$n_0 > 1 $, i.e. it will not hold very close to Anderson's transition.
\par
The end to end resistance is then given by the following sum ${\cal R} =
\sum_{i=0}^N R_i$,
where $N = {L \over r_0} ={2L \over n_0 \xi}$ is the total number of jumps,
and $\ln R_i$ are distributed according
to $\cal P$. The simple addition of resistances amounts to
neglecting resonant tunneling which is justified when
N is large. Let us now distinguish two cases:
\par i) {\it Very long wires}. In this case, the usual central limit theorem
applies and one finds
that the resistance is equal to $N\int_0^\infty dn {\cal P} (n) e^n$. From
this, we find that
$\ln {\overline {\cal R}} \simeq 1.5 n_0$ for $n_0 \sim 1$ (Mott's law), but
that due to the slowly
decreasing tail of
$\cal P$, $\ln {\overline {\cal R}} \simeq {n_0^2 \over 2} \propto T^{-1}$
for $n_0 \gg 1$. This result was first obtained by Kurkijarvi \cite {K}
(see also \cite{Raikh}). \par
ii) {\it Finite wires with $n_0 \gg 1$}. One should however notice that the
largest $n$ drawn from
$\cal P$ in $N$ trials is typically of order $n_{max} \simeq n_0 \sqrt{2\ln
(aN)}$. The central
limit theorem certainly does not apply as long as $n_{max} < \ln {\overline
{\cal R}}$ (the mean cannot
be larger than the largest element !). In this case, the end to end resistance
is entirely governed
by the weakest link, i.e., by $n_{max}$. This can also be understood by
noticing that for intermediate
values of $R=e^n$, the distribution ${\cal P}(R)$ behaves as a power law ${\cal
P}(R) \simeq R^{-(1+\mu)}$
, with an effective exponent $\mu = {\ln R \over 2 n_0^2}$. For $\mu < 1$, the
sum giving $\cal R$ is a
`L\'evy sum' which is well known to be given by its few largest terms
\cite{us}.
 From the above results, one may see that $\mu_{max} = 1$ coincides with the
condition $n_{max} \simeq 2n_0^2$:
the 'anomalous', weak
link dominated regime thus prevails for samples shorter than $N^* \simeq {1
\over a} e^{ 2n_0^2}$, while for
longer wires, one reaches the self-averaging regime.
\par
Experimentally \cite{Sanquer}, $n_0$ is in the range 2-5, $L \simeq 5\mu$m,
$\xi \simeq 20-50$nm, and thus $N$ is between
25 and 100
, which is smaller than $N^*$ at low
temperatures ($T < 1K$)
\footnote{A numerical calculation shows that ${n_{max}\over n_0}$
varies between 2.1 for N=5 to 3.2 for N=100 and 3.8 for N=1000.}
. In
the numerical simulations of P.Lee  \cite{Lee},
 $n_0^2$ varied
between 5 and 40, and $N$ between 12 and 143, again smaller than $N^*$.
Definite theoretical predictions in this regime are thus of great
interest. We have seen that the resistance of the sample is given by $\ln {\cal
R} \simeq n_{max}$. The
average over disorder $\overline {\ln {\cal R}}$ is thus given by: $${\overline
n_{max}} \simeq
n_0 \sqrt{2\ln (aN)}\eqno(3A).$$ We then conclude that for a given $N<N^*$,
$\overline {\ln {\cal R}}$ behaves
as $n_0 \propto T^{-1/2}$, i.e as Mott would predict, but that the {\it slope}
of $\overline {\ln {\cal R}}$
versus $ T^{-1/2}$ increases with N as $\sqrt{2 \ln (aN)}$.
 This result is in agreement with previous investigations
\cite{Serota},\cite{Raikh}
. Let us now turn to the fluctuations: their order of magnitude is governed
by the width of the distribution of $n_{max}$, which can be shown (using the
asymptotic shape of $\cal P$)
to be $\simeq {n_0^2 \over  n_{max}}$.
One then finds: $${\Delta \ln {\cal R} \over \ln {\cal R}} \simeq
{1 \over {2 \ln aN}}\eqno(3B)$$
again
in agreement with \cite{Raikh}
 .
The full distribution of $\ln {\cal R}$
was investigated numerically in \cite{Serota} and analytically in
\cite{Raikh}.{\footnote {This result however relies on the assumption
that the resistances are independent. This is not entirely correct since
two consecutive resistances are correlated through the value $m$ of the
intermediate energy level. }}
\par
It is noteworthy that P.Lee used a 'percolation' method to obtain
the best global conducting path, i.e. the assumption of local optimisation
 is not made. Nevertheless, his numerical results
 \cite{Lee} show all the features  reported above, in particular
 eqs (3A) and (3B). However, if a very detailed analysis of P.Lee's results
is carried out, small differences with our model appear: we performed
 numerical simulations using the same assumptions as P.Lee, except that
we used our local optimisation procedure instead of the percolation
method. We noted two differences:
\par a) First, the percolation method gives values of $\overline
 {\ln {\cal R}}$ slightly lower than predicted by eq.(3A) .
\footnote {To make this comparison, due to differences of notations in \cite
{Lee}, one has to take: $T_0 = 0.04$ and $r_0 \equiv {n_0 \xi \over 4}$.}
However, this difference goes to zero when N increases: for the lowest
temperature and the shortest system ( $N=12$ ) the difference is of 20\%,
 whereas for the highest temperature and the longest system ( $N=143 $)
it is only 12 \%.\par
b) The second point is that the conductance fluctuations
 (see figure 3) as the  Fermi energy is varied are often not symmetric.
 Indeed, as shown in figure 3, one can see that some fluctuations have vertical
slope on one side. We checked that at these points local
optimisation procedure fails, i.e. the optimal path cannot be obtained
by optimising
at each 'step'.
Nevertheless, this does not change much the overall size  of the fluctuations
: for N of order 50, the difference between the
results of ref. \cite{Lee} and  eq.(3B) is
less than 10\%, and goes to zero as N increases.
\par Therefore, we conclude that our assumption of local optimisation is only
exact asymptotically but gives quite good predictions for the sizes
we are interested in.
For the sample studied in ref.\cite{Sanquer}, and in the regime where our model
holds, $\xi$ is estimated to be of the order, or slighly larger than
$l$
, the distance between impurities. This contrasts strongly with P.Lee's
simulations where $\xi$ extends
over 50 localized sites. Since $T_0 \sim \xi ^{-1}$, we expect  much larger
values of $T_0$ for this sample than in ref.\cite{Lee}. \footnote{As a
consequence, the temperature range where Variable Range Hopping holds
should also be much larger than in ref.\cite{Lee}.} Figure 4 shows the
result of a simulation adapted to  sample of ref.\cite{Sanquer}:
${L \over \xi } = 100$, $\xi = 2 l$. Experimental values of $\ln {\cal R}$
are typically $\sim 9$ at $T = 0.45K$ (see the most resistive set of curves
of fig.4 in ref.\cite{Sanquer}). In order to take into account the fact that
our local optimisation method overestimates $\ln {\cal R}$ by $\sim 20\% $,
 we adjusted $T_0$ value so that $\ln {\cal R}$ (as given by (3A)) $
 \simeq 11.5 $ at $T = 0.45K$ and found $T_0 = 6.0K$. $T_0$ is found to
vary between $2K$ and $10K$ from the barely insulating to the strongest
insulating regime observed.
\par
We shall now discuss in more details the nature of these fluctuations {\it for
a given sample},
as the temperature or the gate voltage (i.e. the Fermi level) is varied. \par
{\it i) Temperature.} The main point of the above analysis
is that the resistance of the sample is entirely governed by a certain pair
of impurities, between which transport is most difficult
\cite{Lee},\cite{Raikh}. Crudely speaking, the large resistance arises
either because the distance between the sites is large $\simeq n_{max} \xi$,
but their energies are rather
small (Type I), or because they are close in space but require a large energy
$\simeq n_{max}kT$
 from the thermal bath (Type II). Suppose at a given temperature $T_1$, the
weak link is of type II. Its
resistance thus quickly increases as T decreases. It becomes more and more
probable that this link
will be `shunted' by a longer but less energetic type I link. It is relatively
easy to show, using
excluded area arguments much as above, that this will occur at a temperature
$T_2$ such that, typically
${(T_1 - T_2)\over T_2} \simeq {1 \over \ln (aN)}$. Conversely, a low energy
type I link will soon loose
its `weak link status' as the second largest resistance grows. The temperature
at which the switch occurs is given by the same expression. The temperature
dependence of a given
sample is summarized in figure 2: at low temperatures, it is given by a
succession of plateaus and
activated regions, oscillating around the mean (Mott) behaviour $n_0 \sqrt{2\ln
(aN)}$. The size of
these regions is roughly constant in $\ln$ scale, and goes to zero with the
length of the
sample (as $1\over \ln(aN)$). \par
{\it ii) Gate Voltage}. Following the discussion of P. Lee \cite{Lee}
(see also \cite{Raikh}), one sees from eq. (1)
that as the Fermi energy $E_F$ is varied, two cases may occur: either both
sites are on the
same side of the Fermi level, and $\ln {\cal R}$ varies as $\pm {E_F \over
kT}$, or they are
on opposite sides and $\ln {\cal R}$ is constant. Since the fluctuations of
$\ln {\cal R}$ are
of order  $n_0^2  \over n_{max}$ and assuming that varying $E_F$ is
tantamount to changing the disorder, the peak to peak distance $\Delta E_f$ is
found to be of the order of
${n_0^2 / n_{max} \over 1/2kT} \simeq { 2 n_0 kT \over \sqrt{2 \ln{aN}}}
\simeq { 0.5-1.0 K}$ for the experimental data. This value is, in fact, twice
 the mean level spacing in energy within a box of length $r_{max}$. But, as
very recent simulations at $T = 0 K$ \cite{AvPi} showed, the typical width
in energy of {\it quantum fluctuations} is given by the mean level spacing
within the considered coherent system: intuitively, the transmission
changes
abruptly around every resonance.
Assuming, as usual, that quantum coherence is preserved on the scale of
each hop, we find that quantum fluctuations {\it within the dominant link}
must be taken into account since their typical energy scale is also the mean
level spacing within the box of length $r_{max}$. Indeed, since the resistance
 of a given sample is that of the most resistive link, quantum fluctuations
could only be neglected if they varied more slowly with energy than the
'geometrical' fluctuations, which, {\it typically}, is not true. Of course,
these arguments deal only with average energy scales. Therefore, the precise
origin ( geometrical or quantum ) of fluctuations versus Fermi energy will
depend on each particular c
ase. Experimentally, if an observed fluctuation is strongly temperature
dependent, one will say that this fluctuation is of geometrical origin. Indeed,
if a fluctuation is only due to moving Fermi energy in eq.(1), one see easily
that shifting temperature
 will change it strongly (see figure 1 of \cite{Lee}, and figure 3). On the
contrary, a fluctuation of quantum origin will be temperature independent, on
the whole range in temperature where the dominant link remains the same
\cite{Sanquer}.
\par
 Now, let us consider the case of a fluctuation purely of quantum origin. One
should realize that this fluctuation cannot develop fully, for reasons similar
to
those mentioned above: imagine that as the Fermi energy is varied, one
encounters a resonance which
considerably enhances the transmission coefficient across the weak link. In
this case, this link will
simply disappear from the game, and the second largest resistance will become
the largest (the
probability for simultaneous resonances being negligible). One can show that
the difference between
the $\ln$ of
the largest and the second largest resistance is also of order $n_0 ^2 \over
n_{max}$. Similarly, if
interference effects cause the transmission across the `weak link' to be
considerably less than
its most probable value, a better site will be found, again limiting the drop
to $n_0^2 \over n_{max}$.
We thus conclude that one directly observes (due to the dominance of a single
link) quantum fluctuations
only when they are smaller than  $n_0^2 \over n_{max}$. For larger
fluctuations, these quantum fluctuations
are self consistently truncated because of the presence of the other links. In
particular, contrarily
to the ideas of \cite{Azbel}, one never observes true resonances except at very
low temperatures or
for short samples (see below).
\par
We have thus reached the conclusion that 'truncated quantum fluctuations'
occur when $w_{geo} \gg w_{q} > {n_0^2 \over n_{max}}$. As mentionned
above, one may estimate quantum fluctuations as  :
$w_q = (\Delta \ln {\cal R})_{Quant.} \simeq C_\omega ({r \over \xi})^\omega
\simeq n_{max}^\omega$, with $\omega={1\over 2}$
in the barely localised regime \cite{Pichard} or $\omega \sim {1 \over 5}$ in
the directed path regime
. Comparing with
$n_0^2 \over n_{max}$, one finds that quantum fluctuations are truncated for
long samples or sufficiently high temperatures,
more precisely when $\ln (aN) > n_0^{1 - \omega \over 1 + \omega}$. (For
simplicity, we set $C_\omega= 1$). Both for $\omega= {1 \over 5}$ or $ {1 \over
2}$,
we find that quantum fluctuations are indeed
truncated in our experiments (except perhaps for the lowest temperatures) and
we expect eq. (3B) to hold.
This prediction is compared with the experiments in figure 5, where  $\ln {\cal
R}$ is plotted versus
the smoothed  $\overline{\ln {\cal R}}$
(we took $2 \ln(aN) = 9$). Agreement is reasonnably good. Note in particular
that the experimental sample is
longer than those used in the numerical simulations: this is why  (see
eq.(3B)) the observed fluctuations are
substantially smaller than those reported by P. Lee (fig. 1 of ref.\cite{Lee}).
Finally, the fluctuations
seem to depart somewhat from eq.(3B) for the highest resistances. This may be a
sign that one
enters the purely quantum regime, with a small value of $\omega$.
\par
In conclusion, we have shown, through a detailed analysis of the Mott
conduction mechanism in one
dimension, that the large conductance fluctuations in disordered insulators
result from a subtle
interplay between purely quantum phenomena and geometrical fluctuations arising
from the energies and locations
of the impurities. Our results are in agreement with previous
investigations and are successfully compared to both the
experiments (see the companion paper)
and to numerical simulations.\par
\vskip 1cm
{\bf Acknowledgments}\par
We want to thank J.L. Pichard for many theoretical discussions and M. Sanquer
for having performed the experiments motivating this work and for
constant collaboration. Finally, we thank Pr. B. Shklovskii for drawing our
attention on ref.\cite{Raikh}, which, as we discovered,  contains many of the
ideas
expressed here.\par
\vskip 1cm
{\bf Figure Captions}\par
Figure 1. Starting from a site with energy $m kT$, no sites must be present in
the hatched region of
the $(x,E)$ plane for the resistance to be larger than $e^n$. Insert: Numerical
determination of
${\cal P}(y)$ versus $y = {n \over n_0}$. The most probable value of $y$ is
1.3, and the average of $y$
is 1.5. \par
Figure 2: Evolution of $\ln \ln {\cal R}$ with $-\ln T$ (schematic). For
sufficiently long samples, one
should see three regions: A: Mott, with slope $1/2$, B: Self-averaging regime,
with slope $1$, and
finally weak link dominated region, with a succession of plateaus (slope $0$)
and activated (slope $1$) regions,
around an effective Mott law (dotted line). \par
Figure 3: Simulation of geometrical fluctuations versus Fermi energy,
using the same parameters as in \cite{Lee}: ${L \over \xi } = 1000$,
$\xi = 50 l$, density of states = 1 per Kelvin. Comparison with \cite{Lee}
reveals differences between the shape of fluctuations obtained by the
percolation method and by local optimisation. Note that the vertical slopes
- which are responsible for the  crossings
of curves at  different temperatures- are artefacts of the local
optimisation procedure.
 \par
Figure 4: Simulation of the wire of \cite{Sanquer} with our model.
${L \over \xi } = 100$, $\xi = 2 l$, $T_0 = 6.0$ Kelvins. Predictions of
our theoretical model are in good agreement with numerical results.
 '$<$Correlation energy$>$' means the average energy width of fluctuations.\par
Figure 5:   $\ln {\cal R}$ versus
$\overline{\ln {\cal R}}$ as experimentally determined (see
\cite{Sanquer}); the dotted line shows our prediction for the typical
fluctuations. \par


\begin{thebibliography}{99}
\bibitem{Revue} 'Mesoscopic Phenomena in Solids', B.L. Altshuler, P.A.
Lee and R.A. Webb editors, Elsevier Science Publisher, 1991.
\bibitem{IBM} A. B. Fowler, A. Hartstein, R.A. Webb, Phys. Rev. Lett. 48, 196
(1982), E.I.Laiko,A.O.Orlov,A.K.Savchenko,E.A.Ilyichev and E.A.Poltoratsky Zh
Eksp Teor Fiz. 93, 2204 (1987)(Sov. Phys. JETP 66,1264 (1987)).
\bibitem{Sanquer} F. Ladieu, D. Mailly, M. Sanquer,'Conductance Statistics in
Small
GaAs:Si wires at low temperatures, II: Experimental aspects', companion
paper.
\bibitem{Azbel} M.Y. Azbel, A. Hartstein, D.P. DiVincenzo, Phys. Rev. Lett.52,
1641 (1984), M. Y.
Azbel, Solid State Commum., 45, 527 (1983)
\bibitem{Pichard} \par
J.L. Pichard, N. Zanon, Y. Imry, A.D. Stone, J. Phys. France 51 (1990) 587
\par
J.L. Pichard, in ``Quantum coherence in mesoscopic systems'',B. Kramer Ed.,
NATO Advanced Studies Institute Series (Plenum, New York) to appear (1990).
\par
\bibitem{DP} M. Kardar, E. Madina, Y. Shapir, W. S. Wang, Phys. Rev. Lett., E.
Medina
, PhD Dissertation,June 1991, (MIT), unpublished
\bibitem {Mott} N. F. Mott, (1972), Phil. Mag. 26, 1015
\bibitem{Pollak} N.F.Mott,(1968) J. non-crystalline Solids 1,1
\bibitem{Lee}
P. A. Lee, Phys. Rev. Lett. 53, 2042 (1984)
\bibitem{Serota} R.A. Serota, R.K. Kalia, P.A. Lee, Phys. Rev.B 33
(1986) 8441.
\bibitem{Raikh} M.E Raikh, I.M Ruzin, Sov. Phys. JETP, 68 (1989) 642.
See also M.E. Raikh, I.M. Ruzin, \cite{Revue}, p. 315.
\bibitem{Halp}V. Ambegaokar, B. Halperin, J.S. Langer, Phys. Rev. B4, 2612,
(1971)
\bibitem{Imry}U. Sivan, Y. Imry, Phys. Rev . B 35, 6074, (1987)
\bibitem{K} J. Kurkijarvi, Phys. Rev. B 8, 922 (1973)
\bibitem{us} see, e.g. J.P. Bouchaud, A. Georges, Phys. Rep. 195, 127 (1990),
and
Appendix 1 of J. Physique II 1 (1991) 1465.
\bibitem{AvPi} Avishai, Pichard, Muttalib, companion paper
\end{thebibliography}
\end{document}